# One-Loop Electroweak Radiative Corrections to Bhabha Scattering in the Belle II Experiment

**A. G. Aleksejevs[a], S. G. Barkanova[a], Yu. M. Bystritskiy[b, *], and V. A. Zykunov[b, c]**

[a]*Memorial University, Corner Brook (NL), A2H 5G4 Canada*
[b]*Joint Institute for Nuclear Research, Dubna, 141980 Russia*
[c]*Francisk Skorina Gomel State University, Gomel, 246019 Belarus*
**e-mail: yury.bystritskiy@gmail.com*

**Abstract**—The Standard Model radiative corrections to the Bhabha scattering process are considered within the one-loop approximation. Both virtual corrections and corrections for the real photon emission are taken into consideration. The calculation was performed at the energy assumed at the Belle II (Japan) facility.

The processes of electron–positron pair annihilation into a pair of fermions has been playing a fundamental role in the study of the microcosmos nature since the QED was established to the present time, when the Standard Model of Electroweak Interactions (SM) has obtained the status of an experimentally validated theory. The process of electron-positron annihilation is far from having exhausted its potential so far. The VEPP (Novosibirsk) and BEPC-II (Beijing, China) facilities successfully continue their operation aimed to study the physics of mesons. The program for $B$ meson production at Belle II (KEK, Japan) is of interest. One of the main goals of this program is the detection of the sine of the Weinberg angle, $\sin\theta_W$, the Standard Model (SM) major parameter in the channel $e^-e^+ \to \mu^-\mu^+$ at the energy below the $Z$ resonance [for the Belle II, the energy of reaction $\sqrt{s} = m(\Upsilon_{4S}) = 10.577$ GeV in the $e^-$ and $e^+$ center of mass system (cms)].

The electron-electron (Møller) mode can be used at the ILC/CLIC/FCC electron-positron colliders of the new generation. This mode can be extremely interesting, like those used at lower energies in the E-158 (SLAC) experiment and planed in the MOLLER experiment at JLab both in precision tests and SM measurements, and for a search of new physics (NP) [1].

We set the goal to obtain asymptotic formulas in a simple compact form, but nevertheless reliable (this is proved by successful comparison with the exact results obtained by the FeynArts/FormCalc computer algebra methods [2]) at the energies below the $Z$ resonance.

## 1. BORN APPROXIMATION

The Bhabha scattering process within the SM limits can be written as follows:

$$e^-(p_1) + e^+(p_2) \to e^-(p_3) + e^+(p_4) \quad (1)$$

and is described at the Born level by the diagrams presented in Fig. 1. Four-momenta of the initial particles ($p_1$ and $p_2$) and final particles ($p_3$ and $p_4$) form the standard set of the Lorentz-invariant Mandelstam variables: $s = q_s^2 = (p_1 + p_2)^2$, $t = q_t^2 = (p_1 - p_3)^2$, and $u = (p_2 - p_3)^2$. Below, unless otherwise stated, only the results corresponding to the ultrarelativistic approximation are given: $s, -t, -u \gg m^2$, where $m$ is the mass of electron. The following amplitudes correspond to the diagrams in Fig. 1:

$$\mathcal{M}_t^a = e^2 Q_e^2 D_a(q_t) \times \left[\bar{u}(p_3)\gamma_\mu\Gamma^a u(p_1)\right]\left[\bar{u}(-p_2)\gamma_\mu\Gamma^a u(-p_4)\right], \quad (2)$$

$$\mathcal{M}_s^a = -e^2 Q_e^2 D_a(q_s) \times \left[\bar{u}(-p_2)\gamma_\mu\Gamma^a u(p_1)\right]\left[\bar{u}(p_3)\gamma_\mu\Gamma^a u(-p_4)\right], \quad (3)$$

which in total provide the full amplitude of the process in the Born approximation: $\mathcal{M}_0^a = \mathcal{M}_t^a + \mathcal{M}_s^a$. Here we use the following denotations:

$$D_a(q) = \frac{1}{q^2 - m_a^2}, \quad \Gamma^a = v^a - a^a\gamma_5, \quad a = \gamma, Z. \quad (4)$$

$$v^\gamma = -Q_e, \quad a^\gamma = 0,$$
$$v^Z = \frac{I_e^3 - 2Q_e s_W^2}{2 s_W c_W}, \quad a^Z = \frac{I_e^3}{2 s_W c_W}. \quad (5)$$

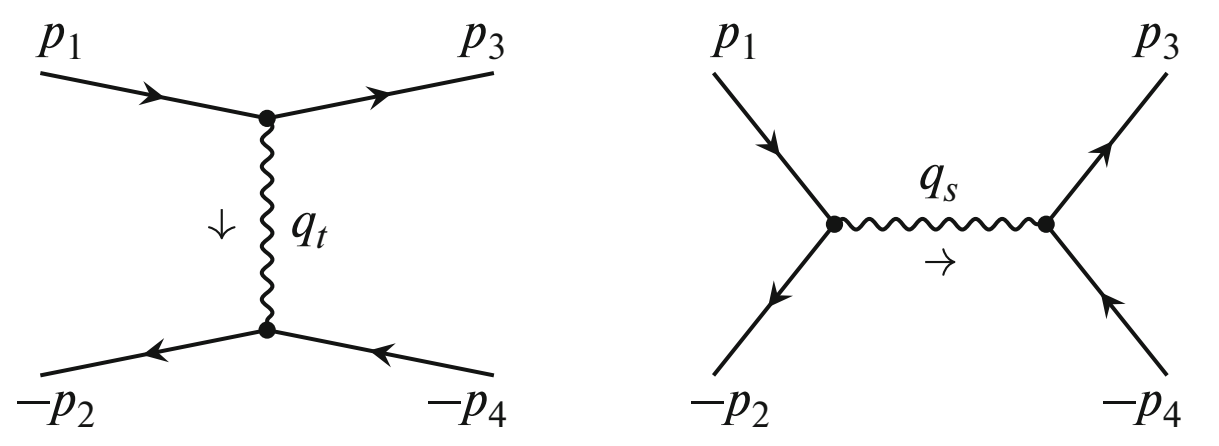

**Fig. 1.** Born approximation.

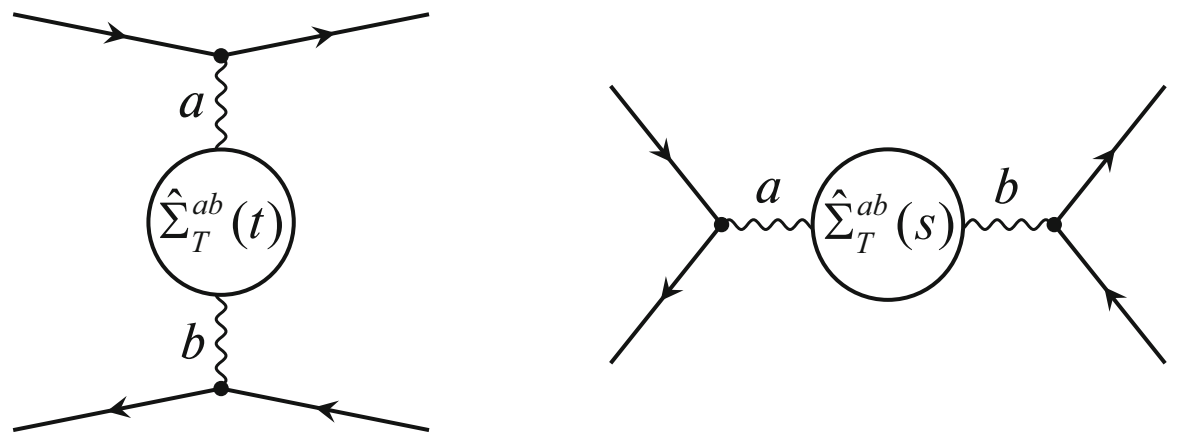

**Fig. 2.** Boson self-energies.

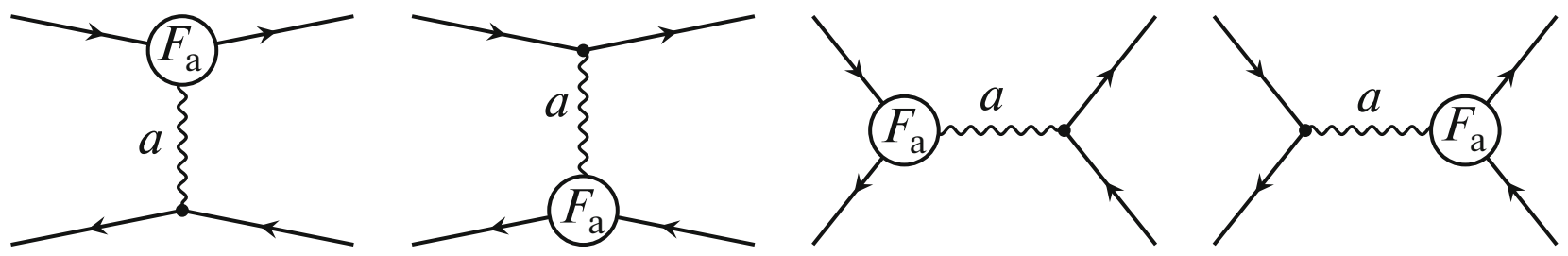

**Fig. 3.** Vertex corrections.

The corresponding cross section can be written in the following form:

$$\frac{d\sigma_0}{dC_{13}} = \frac{\pi\alpha^2}{2s}\sum_{rr'}\sum_{a,b=\gamma,Z}\Pi_{rr'}^{ab}S_{rr'}^{aab}, \tag{6}$$
$$\Pi_{rr'}^{ab} = D_a(q_r)D_b^*(q_{r'}),$$

where the index $r = s,t$ determines the reaction channel, $C_{13} = \cos\theta_{13}$, and $\theta_{13}$ is the scattering angle, i.e., the angle between the vectors $\mathbf{p}_1$ and $\mathbf{p}_3$ in the center of mass system. The quantities $S_{rr'}^{abc}$ can be written as follows:

$$\begin{aligned} S_{tt}^{abc} &= \mathrm{Sp}\left[\gamma_\mu\Gamma^a U_1\Gamma^{c+}\gamma_\nu U_3\right]\mathrm{Sp}\left[\gamma_\mu\Gamma^b U_4\Gamma^{c+}\gamma_\nu U_2\right], \\ S_{ts}^{abc} &= -\mathrm{Sp}\left[\gamma_\mu\Gamma^a U_1\Gamma^{c+}\gamma_\nu U_2\gamma_\mu\Gamma^b U_4\Gamma^{c+}\gamma_\nu U_3\right], \\ S_{st}^{abc} &= -\mathrm{Sp}\left[\gamma_\mu\Gamma^a U_1\Gamma^{c+}\gamma_\nu U_3\gamma_\mu\Gamma^b U_4\Gamma^{c+}\gamma_\nu U_2\right], \\ S_{ss}^{abc} &= \mathrm{Sp}\left[\gamma_\mu\Gamma^a U_1\Gamma^{c+}\gamma_\nu U_2\right]\mathrm{Sp}\left[\gamma_\mu\Gamma^b U_4\Gamma^{c+}\gamma_\nu U_3\right], \end{aligned} \tag{7}$$

where $U_i$ are the spin density matrices:

$$U_1 \equiv u(p_1)\bar{u}(p_1) = \frac{1}{2}(1+\lambda_1\gamma_5)(\hat{p}_1 + m),$$
$$U_2 \equiv u(-p_2)\bar{u}(-p_2) = \frac{1}{2}(1-\lambda_2\gamma_5)(\hat{p}_2 - m).$$
$$U_3 \equiv u(p_3)\bar{u}(p_3) = \hat{p}_3 + m,$$
$$U_4 \equiv u(-p_4)\bar{u}(-p_4) = \hat{p}_4 - m.$$

The $\gamma$ matrix traces (7) can be expressed in terms of the following combinations of the initial particle polarizations $\lambda_{1,2}$:

$$P_1^\pm = \lambda_1 \pm \lambda_2, \quad P_2^\pm = 1 \pm \lambda_1\lambda_2,$$

and the combinations of the coupling constants:

$$f_\pm^{abc} = g_V^{ac}g_V^{bc} \pm g_A^{ac}g_A^{bc}, \quad g_\pm^{abc} = g_V^{ac}g_A^{bc} \pm g_A^{ac}g_V^{bc},$$
$$g_V^{ab} = v^a v^b + a^a a^b, \quad g_A^{ab} = v^a a^b + a^a v^b,$$

in the following form:

$$\frac{1}{2}S_{tt}^{abc} = P_2^- f_+^{abc}u^2 + P_2^+ f_-^{abc}s^2 - P_1^- g_+^{abc}u^2 + P_1^+ g_-^{abc}s^2,$$
$$\frac{1}{2}S_{ts}^{abc} = \frac{1}{2}S_{st}^{abc}P_2^- f_+^{abc}u^2 - P_1^- g_+^{abc}u^2, \tag{8}$$
$$\frac{1}{2}S_{ss}^{abc} = P_2^-\left(f_+^{abc}u^2 + f_-^{abc}t^2\right) - P_1^-\left(g_+^{abc}u^2 - g_-^{abc}t^2\right).$$

Let us introduce the relative corrections $\delta_\pm$:

$$\delta_+^C = \frac{\sigma_{LL}^C + \sigma_{LR}^C + \sigma_{RL}^C + \sigma_{RR}^C}{\sigma_{LL}^0 + \sigma_{LR}^0 + \sigma_{RL}^0 + \sigma_{RR}^0} = \frac{\sigma_{00}^C}{\sigma_{00}^0}, \tag{9}$$

$$\delta_-^C = \frac{(\sigma_{LL}^C + \sigma_{LR}^C) - (\sigma_{RL}^C + \sigma_{RR}^C)}{(\sigma_{LL}^0 + \sigma_{LR}^0) - (\sigma_{RL}^0 + \sigma_{RR}^0)} = \frac{\sigma_{L0}^C - \sigma_{R0}^C}{\sigma_{L0}^0 - \sigma_{R0}^0}, \tag{10}$$

where the subscripts $\{\lambda_1\lambda_2\}$ determine the initial particles polarizations, and the superscript $C$ sets the type of contribution.

## 2. RADIATIVE CORRECTIONS

The radiative corrections consist of contributions of several types: boson self-energies (Fig. 2), vertex corrections (Fig. 3), exchange of two virtual bosons (Fig. 4), and corrections for real photon emission (Fig. 5).

The boson self-energies (Fig. 2) can be written in the form similar to the Born form:

$$\frac{d\sigma_{\mathrm{BSE}}}{dC_{13}} = \frac{\pi\alpha^2}{s}\sum_{rr'}\sum_{a,b,c=\gamma,Z}\Pi_{rr'}^{abc}S_{rr'}^{abc}, \tag{11}$$

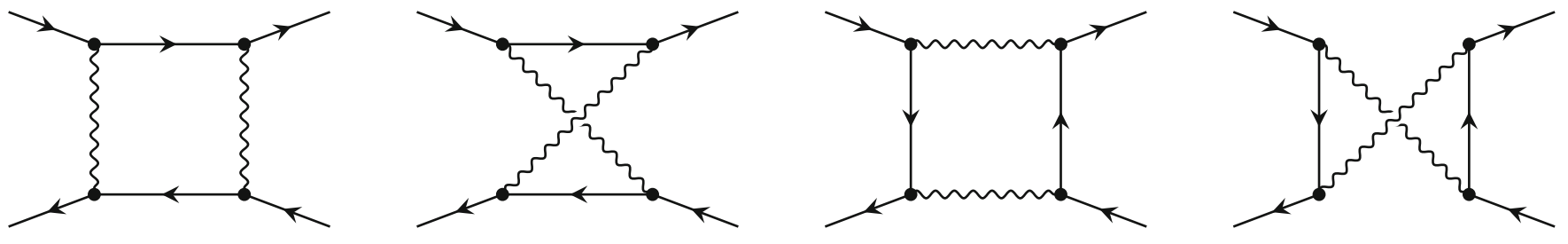

**Fig. 4.** Two-boson corrections (boxes).

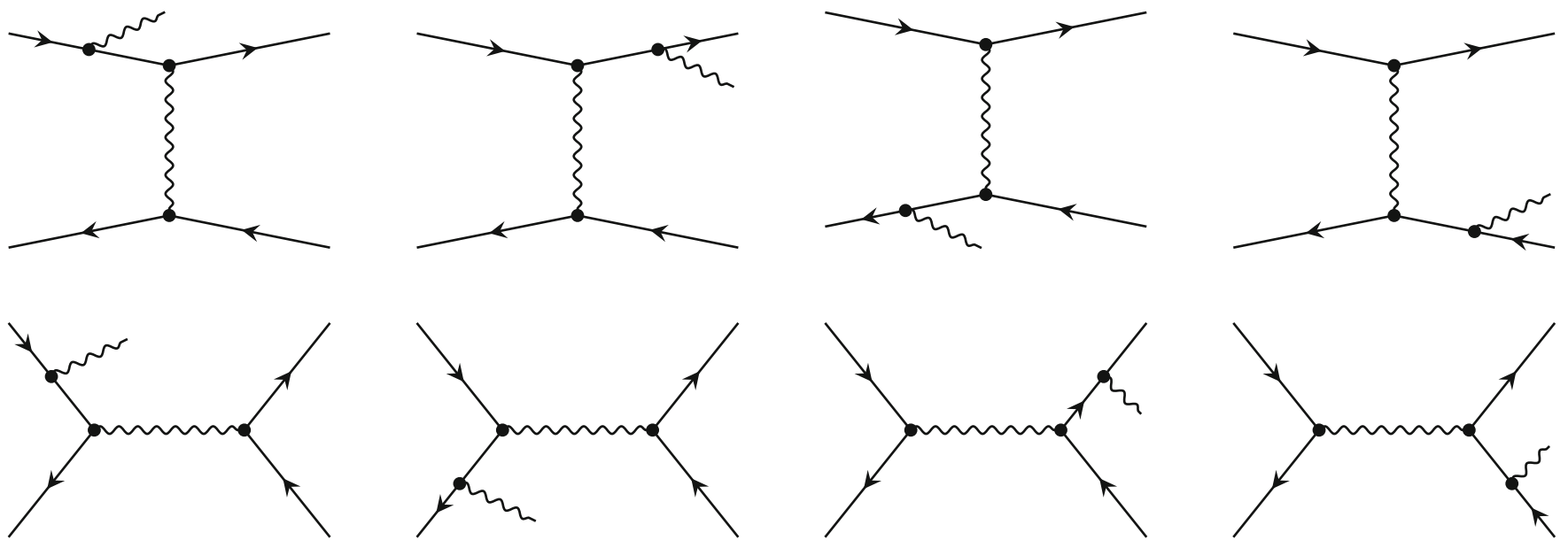

**Fig. 5.** Corrections for the photon emission.

where the transverse part of the self-energy operator $\hat{\Sigma}^{ab}_T(r)$ is included in the following combination:

$$\Pi^{abc}_{rr'} = -D_a(q_r)\hat{\Sigma}^{ab}_T(r)D_b(q_r)D^*_c(q_{r'}). \quad (12)$$

In the vertex corrections (Fig. 3), there are already two Born-like contributions corresponding to the modifications of one or another vertex from the Born diagram (Fig. 1):

$$\frac{d\sigma_{\text{Ver}}}{dC_{13}} = \frac{\pi\alpha^2}{s}\sum_{rr'}\sum_{a,b=\gamma,Z}\Pi^{ab}_k\left(S^{F_a ab}_{rr'} + S^{aF_a b}_{rr'}\right), \quad (13)$$

where the summation is performed over the channels $r,r' = s,t$, and the traces of the $\gamma$ matrices $S^{abc}_{rr'}$ coincide with the Born ones (8), only the trivial vertices $v^a$ and $a^a$ are replaced by the renormalized vertex form factors:

$$v^a \to v^{F_a},\quad a^a \to a^{F_a},\quad \text{where}\quad a = \gamma, Z, \quad (14)$$

and where:

$$v^{F_\gamma} = \frac{\alpha}{4\pi}\left(v^\gamma\Lambda^\gamma_1 + \left((v^Z)^2 + (a^Z)^2\right)\Lambda^Z_2 + \frac{3}{4s^2_W}\Lambda^W_3\right), \quad (15)$$

$$a^{F_\gamma} = \frac{\alpha}{4\pi}\left(a^\gamma\Lambda^\gamma_1 + 2v^Z a^Z\Lambda^Z_2 + \frac{3}{4s^2_W}\Lambda^W_3\right), \quad (16)$$

$$v^{F_Z} = \frac{\alpha}{4\pi}\left(v^Z\Lambda^\gamma_1 + v^Z\left((v^Z)^2 + 3(a^Z)^2\right)\Lambda^Z_2 + \frac{1}{8s^3_W c_W}\Lambda^W_2 - \frac{3c_W}{4s^3_W}\Lambda^W_3\right), \quad (17)$$

$$a^{F_Z} = \frac{\alpha}{4\pi}\left(a^Z\Lambda^\gamma_1 + a^Z\left(3(v^Z)^2 + (a^Z)^2\right)\Lambda^Z_2 + \frac{1}{8s^3_W c_W}\Lambda^W_2 - \frac{3c_W}{4s^3_W}\Lambda^W_3\right). \quad (18)$$

The values of $\Lambda_i$ are well known [3]. The two-boson exchanges (box-type diagrams) (Fig. 4) contain the exchange of different bosons:

$$\mathcal{M}_{\text{Box}} = \mathcal{M}^{\gamma\gamma}_t + \mathcal{M}^{\gamma Z}_t + \mathcal{M}^{Z\gamma}_t + \mathcal{M}^{ZZ}_t + \mathcal{M}^{WW}_t + (t \to s). \quad (19)$$

We denote straight boxes as $\mathcal{M}^{ab}_{r,D}$, and the boxes with crossed boson propagators as $\mathcal{M}^{ab}_{r,C}$ and write:

$$\mathcal{M}_{\text{Box}} = \sum_{a,b}\left(\mathcal{M}^{ab}_{t,D} + \mathcal{M}^{ab}_{t,C} + \mathcal{M}^{ab}_{s,D} + \mathcal{M}^{ab}_{s,C}\right). \quad (20)$$

The corresponding contribution to the cross section can be written in the following form:

$$\frac{d\sigma_{\text{Box}}}{dC_{13}} = \frac{1}{2^4\pi s}\text{Re}\sum_{c=\gamma,Z}\mathcal{M}_{\text{Box}}\left(\mathcal{M}^c_0\right)^+.$$

As an example, we consider a contribution from the following combination in $\mathcal{M}_{\text{Box}}\left(\mathcal{M}^c_0\right)^+$:

$$\mathcal{M}^{ab}_{t,D}\left(\mathcal{M}^c_t\right)^+ = 4\pi\alpha^3\int\frac{dk}{i\pi^2}D_a(k)D_b(q_t - k)D^*_c(q_t) \times \text{Sp}\left[\gamma_\beta\Gamma^b S(p_1 - k)\gamma_\mu U^a_1\Gamma^{c+}\gamma_\nu U_3\right] \times \text{Sp}\left[\gamma_\mu\Gamma^a S(-p_2 - k)\gamma_\beta\Gamma^b U_4\gamma_\nu U^c_2\right]. \quad (21)$$

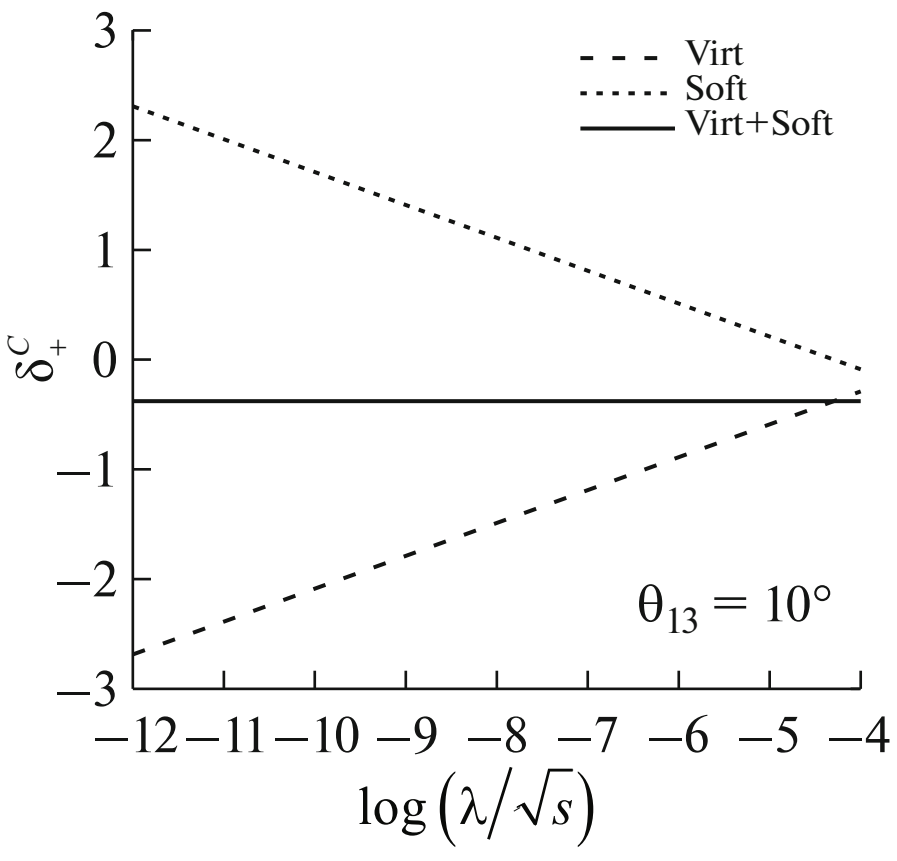

**Fig. 6.** Cancellation of the infrared divergence.

One of the most complicated contributions of such type is the case of two-photon exchange ($\gamma\gamma$ box). Within our approach, it has the following form:

$$\mathcal{M}^{\gamma\gamma}_{t,D}\left(\mathcal{M}^c_t\right)^+ = \frac{8\pi\alpha^3}{t}D_c^*(q_t)(P_3x_1 + P_4y_1),$$
$$x_1 = (L_t - L_s)(S_1(L_s - L_t) - 2tu) + 2u^2\left(L_s(L_s - 2L_\lambda) - \frac{4}{3}\pi^2\right),$$
$$y_1 = 2s^2\left(L_t^2 - 2L_s(L_t - L_\lambda) + \frac{4}{3}\pi^2\right), \quad (22)$$

where $L_s \equiv \ln\frac{s}{m^2}$, $L_t \equiv \ln\frac{-t}{m^2}$, $L_\lambda \equiv \ln\frac{\lambda^2}{m^2}$, $S_1 \equiv s^2 + u^2$ and the following combinations of helicities and vertices are used:

$$P_3 \equiv P_1^- g_+^{\gamma\gamma c} - P_2^- f_+^{\gamma\gamma c}, \quad P_4 \equiv P_2^+ f_-^{\gamma\gamma c}, \quad P_5 \equiv P_2^- f_-^{\gamma\gamma c}. \quad (23)$$

Apparently, the boxes with photons suffer from the infrared divergence that is regularized by the photon fictional mass $\lambda$. This nonphysical quantity is cancelled in the sum with the contribution of a real photon emission (Fig. 5). The contribution of a soft photon (with the energy $\omega$, such that $\lambda < \omega < \Delta E$, where $\Delta E \ll \sqrt{s}$ is the softness threshold) is usually calculated. Then the matrix element is factorized:

$$\mathcal{M}|_{k\to 0} \sim e_\alpha(k)\left(\frac{p_1^\alpha}{(kp_1)} - \frac{p_2^\alpha}{(kp_2)} - \frac{p_3^\alpha}{(kp_3)} + \frac{p_4^\alpha}{(kp_4)}\right)\mathcal{M}_0^a. \quad (24)$$

The cross section of the real soft photon emission can then be presented in the form:

$$\frac{d\sigma_{\text{soft}}}{dC_{13}} = \delta_{\text{soft}}\frac{d\sigma_0}{dC_{13}}, \quad (25)$$

where

$$\delta_{\text{soft}} = -\frac{\alpha}{4\pi^2} \times \int\limits_{\lambda<\omega<\Delta E}\frac{d^3\mathbf{k}}{\omega}\left(\frac{p_1}{(kp_1)} - \frac{p_2}{(kp_2)} - \frac{p_3}{(kp_3)} + \frac{p_4}{(kp_4)}\right)^2$$
$$= \frac{2\alpha}{\pi}\left(2\ln\frac{2\Delta E}{\lambda}\left(\ln\frac{st}{m^2u} - 1\right) + L_s - \frac{1}{2}L_s^2 - \frac{\pi^2}{3} + \text{Li}_2\left(-\frac{t}{u}\right) - \text{Li}_2\left(-\frac{u}{t}\right)\right). \quad (26)$$

The cancellation of infrared contributions in the sum of virtual (including boxes) and soft corrections (at $\Delta E = 0.01\sqrt{s}$) is presented in Fig. 6.

The emission of a hard photon with the energy $\omega > \Delta E$ can be described by the following exact formula (the kinematics and used notations are presented in Fig. 7):

$$\frac{d\sigma_R}{dC_{13}} = \frac{\alpha^3}{4\pi s}\int\limits_{\Delta E}^{\omega_{\max}}\omega d\omega\int\limits_{-1}^{1}dC_{k1} \times \int\limits_0^{2\pi}d\varphi_k\frac{|\mathbf{p}_3|\theta_{\exp}}{E_4|f'(E_3)|}\sum\left|\mathcal{M}^\gamma\right|^2, \quad (27)$$

where $\theta_{\exp}$ determines all necessary experimental constrains and the detector sensitivity (for example, when the limitation is imposed on the positron outgoing angle $\theta_{24}$, we can assume $\theta_{\exp} = \Theta\left(\pi - \theta_{24}^{\text{cut}} - \theta_{24}\right)\Theta\left(\theta_{24} - \theta_{24}^{\text{cut}}\right)$), and the integration over the photon energy $\omega$ is carried out up to the maximum possible energy of the photon:

$$\omega_{\max} = \frac{\sqrt{s}}{2} - \frac{2m^2}{\sqrt{s}}. \tag{28}$$

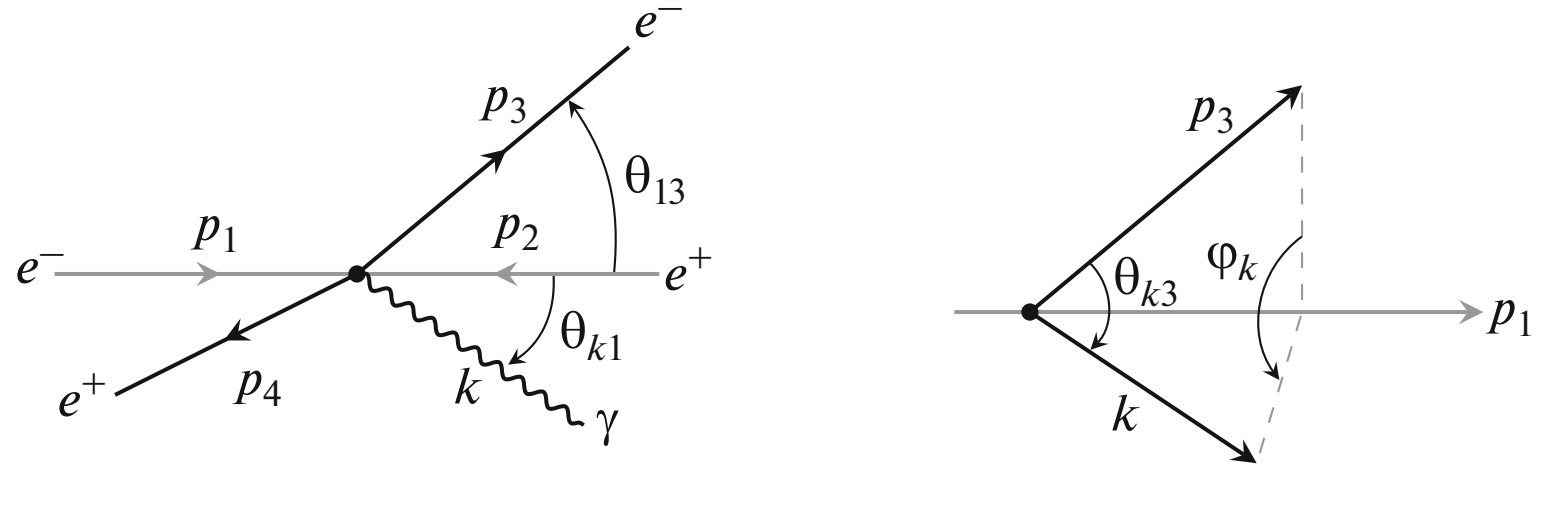

**Fig. 7.** Kinematics of the real photon.

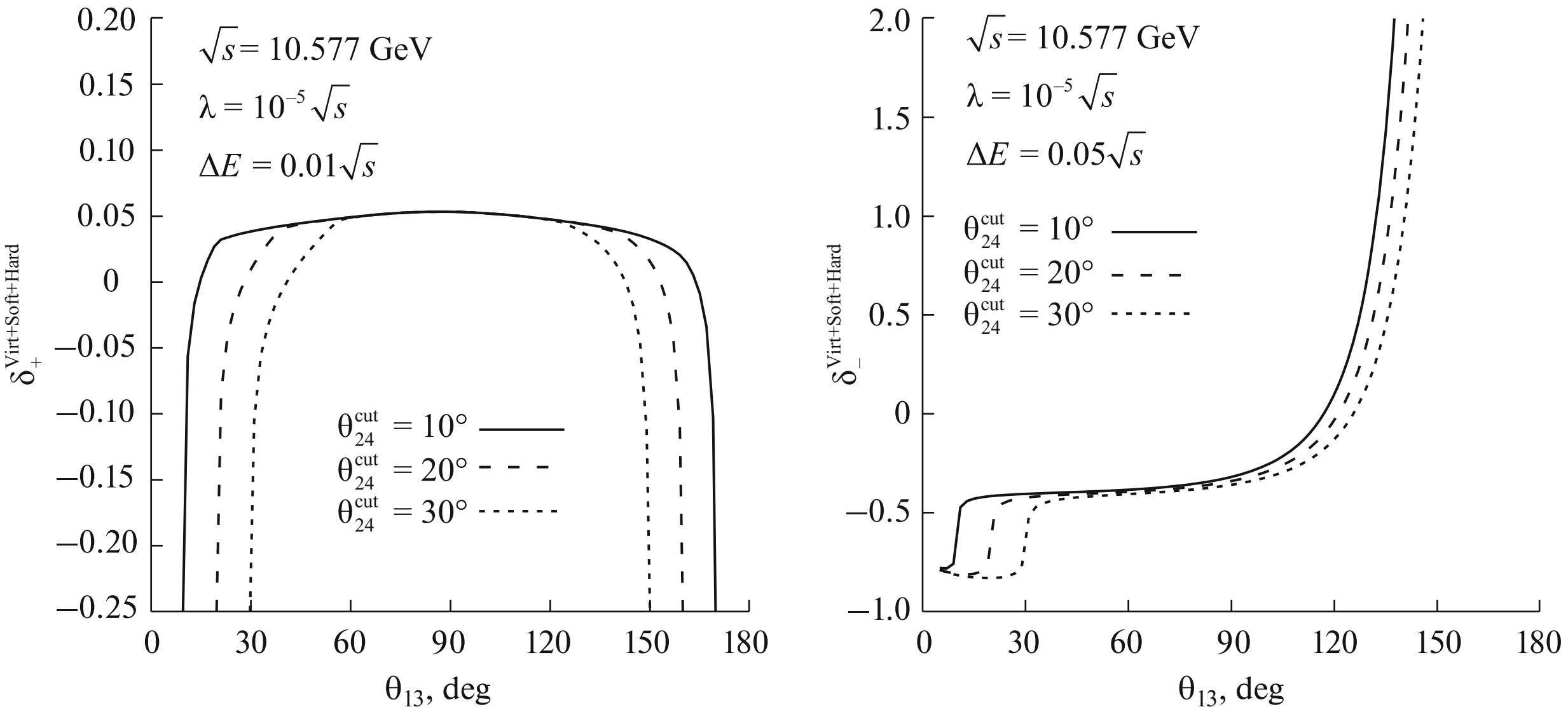

**Fig. 8.** Total contribution of radiative corrections.

The factor $|f'(E_3)|$ in (27) arises due to the careful handling the $\delta$ function that provides the energy conservation. The total contribution to the relative corrections (including the hard real photon emission (27)) is presented in Fig. 8.

## 3. COMPARISON WITH OTHER CALCULATIONS

We compared our asymptotic calculations with the results of the FeynArts/FormCalc automated system of analytical calculations [2] and showed rather good agreement in the region far from the $Z$ boson (see Fig. 9).

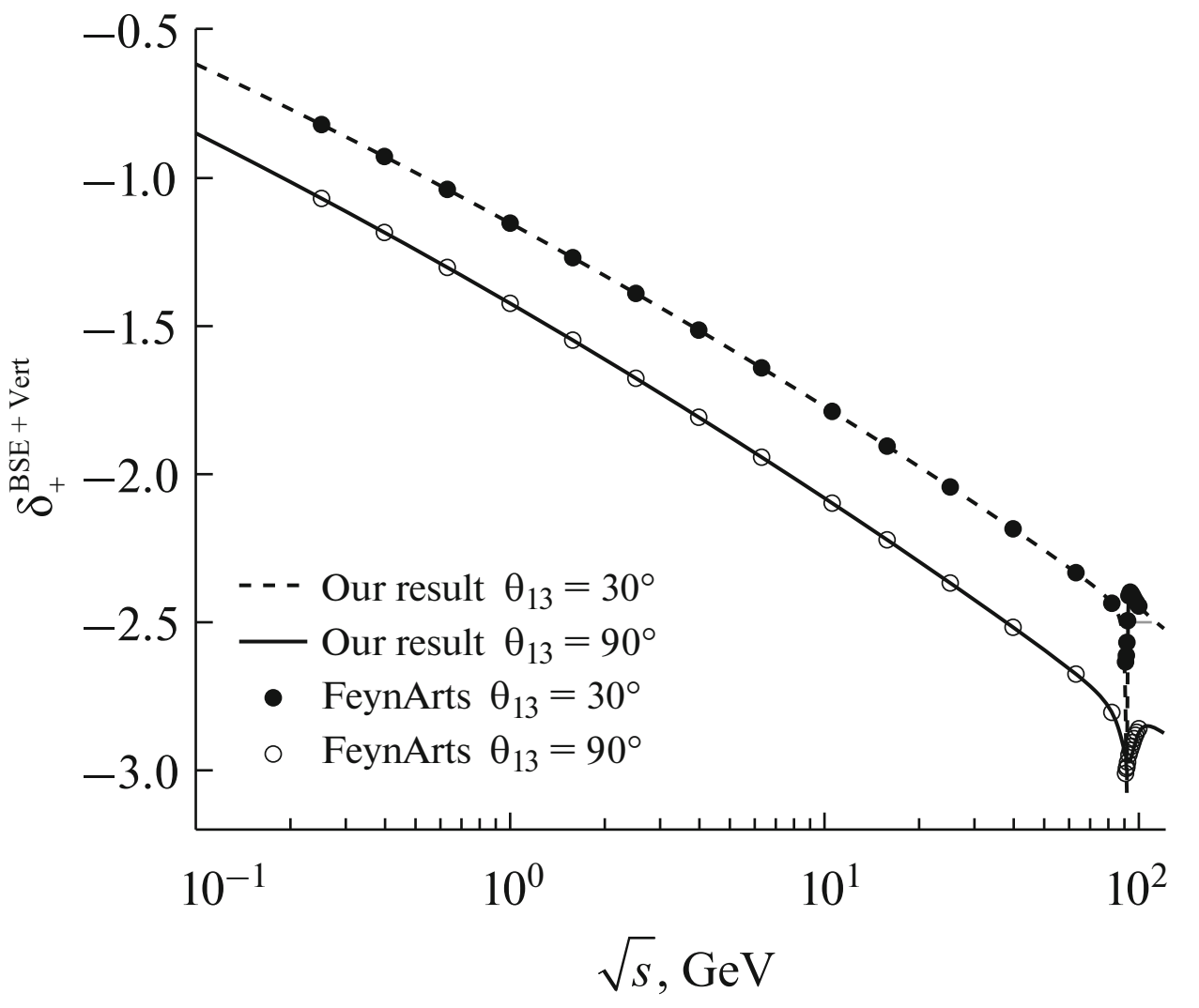

**Fig. 9.** Comparison with the FeynArts/FormCalc.

In addition, we checked the correspondence of our results on the contributions of a real photon emission for different polarizations of the initial particles with the results from SANC and WHIZARD [4]. When the scattering energy was $\sqrt{s} = 500$ GeV, we integrated over the photon energy from the value $\omega = 1$ GeV up to the greatest possible energy $\omega_{\max}$. The results of comparison are presented in Table 1.

**Table 1.** Comparison of the cross section of the hard bremsstrahlung (in pb) with the results from the SANC group [4]

| Group | 00 | LL | RR | LR | RL |
|---|---|---|---|---|---|
| SANC | 15.137(2) | 11.454(3) | 11.455(3) | 20.489(5) | 17.149(4) |
| WHIZARD | 15.138(2) | 11.461(2) | 11.457(2) | 20.488(3) | 17.147(3) |
| Our result | 15.16 ± 0.02 | 11.44 ± 0.02 | 11.44 ± 0.02 | 20.47 ± 0.03 | 17.06 ± 0.02 |

*Translated by N. Semenova*